\documentclass[10pt,conference]{IEEEtran}
\usepackage{cite}
\usepackage{amsmath,amssymb,amsfonts}
\usepackage{algorithmic}
\usepackage{graphicx}
\usepackage{textcomp}
\usepackage{xcolor}
\usepackage[hyphens]{url}
\usepackage{fancyhdr}
\usepackage{hyperref}

\newcommand{\hpcayear}{2026}

\newcommand{\hpcasubmissionnumber}{182}
\title{MoEntwine: Unleashing the Potential of Wafer-scale Chips for Large-scale Expert Parallel Inference}
\usepackage[ruled,linesnumbered, noend]{algorithm2e}
\newcommand{\textbfIt}[1]{\textbf{\textit{#1}}}
\usepackage{float}
\usepackage{booktabs}
\usepackage{makecell}

\usepackage{soul, color, xcolor} 
\soulregister{\ref}{7} 
\soulregister{\cite}{7} 

\newcommand\hpcaauthors{First Author$\dagger$ and Second Author$\ddagger$}
\newcommand\hpcaaffiliation{First Affiliation$\dagger$, Second Affiliation$\ddagger$}
\newcommand\hpcaemail{Email(s)}

\def\hpcacameraready{}

\author{
  \ifdefined\hpcacameraready
    \IEEEauthorblockN{\hpcaauthors{}}
      \IEEEauthorblockA{
        \hpcaaffiliation{} \\
        \hpcaemail{}
      }
  \else
    \IEEEauthorblockN{\normalsize{HPCA \hpcayear{} Submission
      \textbf{\#\hpcasubmissionnumber{}}} \\
      \IEEEauthorblockA{
        Confidential Draft \\
        Do NOT Distribute!!
      }
    }
  \fi 
}

\renewcommand{\hpcaauthors}{
  Xinru Tang\IEEEauthorrefmark{1},
  Jingxiang Hou\IEEEauthorrefmark{1},
  Dingcheng Jiang\IEEEauthorrefmark{1},
  Taiquan Wei\IEEEauthorrefmark{1},
  Jiaxin Liu\IEEEauthorrefmark{1},
  Jinyi Deng\IEEEauthorrefmark{1},\\
  Huizheng Wang\IEEEauthorrefmark{1},
  Qize Yang\IEEEauthorrefmark{1},
  Haoran Shang\IEEEauthorrefmark{1},
  Chao Li\IEEEauthorrefmark{2},
  Yang Hu\IEEEauthorrefmark{1},
  Shouyi Yin\IEEEauthorrefmark{1}\IEEEauthorrefmark{3},
}

\renewcommand{\hpcaaffiliation}{
\IEEEauthorrefmark{1}Tsinghua University, School of Integrated Circuits, BNRist, Beijing, China\\
\IEEEauthorrefmark{2}Shanghai Jiao Tong University, Shanghai, China\\
\IEEEauthorrefmark{3}Shanghai Artificial Intelligence Laboratory, Shanghai, China}

\renewcommand{\hpcaemail}{
  \IEEEauthorrefmark{1}\{tangxr23, houjx22, jdc24, weitq24, jiaxin-l24, wanghz22, yqz23, shanghr23\}@mails.tsinghua.edu.cn, \\
  \IEEEauthorrefmark{1}dengjinyi@mail.tsinghua.edu.cn, hu\_yang@tsinghua.edu.cn, yinsy@tsinghua.edu.cn \\
  \IEEEauthorrefmark{2}lichao@cs.sjtu.edu.cn
}

\begin{document}
\maketitle


\newcommand{\hpcaheight}{0mm}
\ifdefined\eaopen
\renewcommand{\hpcaheight}{12mm}
\fi


\begin{abstract} 

As large language models (LLMs) continue to scale up, mixture-of-experts (MoE) has become a common technology in SOTA models. MoE models rely on expert parallelism (EP) to alleviate memory bottleneck, which introduces all-to-all communication to dispatch and combine tokens across devices. However, in widely-adopted GPU clusters, high-overhead cross-node communication makes all-to-all expensive, hindering the adoption of EP. Recently, wafer-scale chips (WSCs) have emerged as a platform integrating numerous devices on a wafer-sized interposer. WSCs provide a unified high-performance network connecting all devices, presenting a promising potential for hosting MoE models. Yet, their network is restricted to a mesh topology, causing imbalanced communication pressure and performance loss. Moreover, the lack of on-wafer disk leads to high-overhead expert migration on the critical path.

To fully unleash this potential, we first propose Entwined Ring Mapping (ER-Mapping), which co-designs the mapping of attention and MoE layers to balance communication pressure and achieve better performance. We find that under ER-Mapping, the distribution of cold and hot links in the attention and MoE layers is complementary. Therefore, to hide the migration overhead, we propose the Non-invasive Balancer (NI-Balancer), which splits a complete expert migration into multiple steps and alternately utilizes the cold links of both layers. Evaluation shows ER-Mapping achieves communication reduction up to $62\%$. NI-Balancer further delivers $54\%$ and $22\%$ improvements in MoE computation and communication, respectively. Compared with the SOTA NVL72 supernode, the WSC platform delivers an average $39\%$ higher per-device MoE performance owing to its scalability to larger EP.

\end{abstract}

\section{introduction}

\begin{figure}[t]
    \centering
    \includegraphics[width=1.0\linewidth]{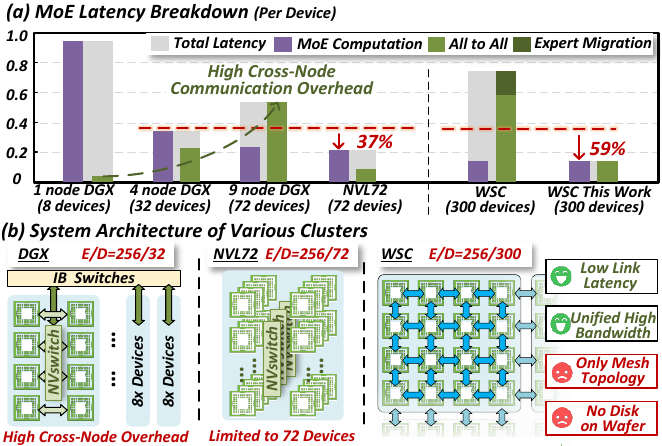}
    \caption{(a) MoE Latency Breakdown of DeepSeek-V3 with $EP$ equals to device count, total latency equals to the maximum of computation and communication time. (b) System Architecture of DGX, NVL72, and WSC.}
    \label{fig:ScaleUp}
\end{figure}


Traditional machine learning systems have reached a mature stage with well-established methodologies and deployments in various domains \cite{Song1, Song2, Song3}, but recent advances in large language models (LLMs) have rapidly surpassed these conventional approaches. LLMs have achieved state-of-the-art performance across a wide range of applications \cite{touvron2023llama,carion2020end,dosovitskiy2021an,wang2025large,song2024tackling}. To further sustain this scaling trend, \textit{Mixture-of-Experts} (MoE) architectures \cite{OutrageouslyMoE} have gained popularity for improving parameter efficiency. Recent models such as DeepSeek-R1-671B \cite{liu2024deepseek} and Qwen3-234B \cite{qwen3report} adopt MoE designs with hundreds of lightweight experts, selectively activating a small subset per token (e.g., 8 out of 256 experts). This design reduces compute cost but imposes a significant memory footprint—especially when multiple experts are colocated on the same device during inference.


To address this, \textit{Expert Parallelism} (EP) \cite{hwang2023tutel,zhai2023smartmoe} distributes experts across devices, ideally one expert per device, to alleviate memory pressure. However, EP requires all-to-all communication to route tokens to and from the activated experts, with overhead scaling rapidly as the device count increases. Thus, a critical factor influencing EP performance is the ratio of experts to available devices, defined as the \textbf{E/D ratio}. A lower E/D ratio indicates fewer experts per device, reducing memory contention and improving inference throughput. The optimal EP performance is theoretically achieved when E/D equals one, provided all devices are interconnected by a unified high-bandwidth, low-latency network.  




However, as shown in Fig. \ref{fig:ScaleUp}(b), in widely-deployed DGX systems \cite{nvidiaDGX}, high-performance networking is confined to each 8-GPU nodes, with high-overhead inter-node links (e.g., IB Link \cite{IBLink}) degrading cross-node communication \cite{li2024locmoe, pan2024parm}. As Fig. \ref{fig:ScaleUp}(a) shows, when the cluster scale exceeds $4$ nodes ($32$ GPUs), the all-to-all overhead exceeds computation by \emph{2.3×}, forcing suboptimal $E/D$=$8$ ($256/32$) and significant performance loss. 
This highlights the need for system-wide high-speed interconnects to minimize E/D and unlock EP scalability.
To address this, NVIDIA introduced the NVL72 supernode \cite{NVL72}, connecting 72 GB200 dies via a custom scale-up network. It improves E/D to 3.6 and boosts performance by 37\% over 4-node DGX. However, its reliance on numerous switches and cables leads to high energy and infrastructure costs, limiting scalability and preventing E/D = 1.


Recently, \textbf{wafer-scale chips (WSCs)} \cite{yu2025cramming,14336-processor,dojo,WSE-3} have emerged as a promising approach to overcome these scaling bottlenecks. By directly interconnecting compute dies via wafer-scale interposers, WSCs offer unprecedented bandwidth and latency characteristics. Tesla’s Dojo platform \cite{dojo}, for instance, achieves 4 TB/s intra-wafer bandwidth between dies and 9 TB/s inter-wafer bandwidth, significantly outperforming NVLink’s 1.8 TB/s (by 4.4×). Such designs facilitate a unified, high-performance network spanning 300 dies in a single cabinet, enabling even E/D ratios below one, which further boosts EP performance up to 59\%. 

Despite their theoretical advantages, directly porting existing GPU-cluster optimizations onto WSCs fails to fully exploit their capabilities due to two unique challenges. First, signal integrity (SI) constraints compel practical WSC implementations \cite{WSE-3, dojo} to adopt mesh topologies instead of ideal all-to-all networks. As a result, all-to-all communication traffic must traverse multiple hops, causing significant congestion and performance degradation in the wafer's central regions. Second, the lack of on-wafer storage exacerbates congestion: expert migration, widely utilized for load balancing, must frequently transfer large expert weights via the already congested wafer interconnects, further degrading performance. 


Motivated by these unique challenges, we introduce \textbf{MoEntwine}, a specialized MoE scaling solution for WSCs, featuring two novel designs: \textbf{Entwined Ring Mapping (ER-Mapping)} and \textbf{Non-invasive Balancer (NI-Balancer)}. 

We first observe that MoE inference workloads involve two primary types of collective communication: the all-to-all operations for token dispatching in MoE layers, and the all-reduce operations within attention layers. Notably, these two types of communication exhibit  different latency characteristics as system scale increases; all-to-all latency escalates quickly with increased device count, while all-reduce latency remains relatively stable. Critically, the parallelism mapping strategy of the attention layers significantly affects the initial token distribution, thus indirectly influencing communication overhead in subsequent MoE layers. To capture this interaction systematically, we propose the \textbf{Full Token Domain (FTD)} framework, analyzing the trade-off between all-to-all and all-reduce overheads. Guided by this analysis, \textbf{ER-Mapping} co-designs the parallelism mapping strategies for attention and MoE layers, balancing communication pressure and dramatically reducing latency. 

Secondly, we surprisingly find that ER-Mapping provides an opportunity to hide the overhead of expert migration. By analyzing the link traffic, it's observed that the distribution of ``hot links'' and ``cold links'' in the attention and MoE layers is complementary. Therefore, we can split a complete expert migration into multiple steps and use the cold links in these two layers alternately without overhead. Based on this, we propose \textbf{NI-Balancer}
, a multi-step expert migration scheme that strategically exploits idle ("cold") communication links in both layers to perform expert weight transfers without additional overhead. Specifically, NI-Balancer first identifies temporal locality patterns of expert selection during inference, then orchestrates expert migration across layers, effectively hiding migration overhead and ensuring agile load balancing. 

Our evaluations show that WSC inherently reduces communication latency by 56\% compared to DGX, benefiting from its unified wafer-scale interconnect. Further optimizations via ER-Mapping achieve up to 62\% additional latency reduction. NI-Balancer completely eliminates expert migration overhead while significantly improving load balance, reducing MoE computation and communication latency by up to 54\% and 22\%, respectively. These innovations effectively address the fundamental communication and migration bottlenecks inherent to wafer-scale EP implementations. Compared to the state-of-the-art NVL72 supernode, the WSC enhanced by MoEntwine achieves 39\% higher average per-device MoE performance, unlocking the full scaling potential of wafer-scale MoE inference.

\section{background}

In this section, we first introduce the structure of LLMs with MoE. We then describe the architecture of wafer-scale chips.

\subsection{Large Language Models with Mixture-of-Experts}

\begin{figure}[t]
    \centering
    \includegraphics[width=1.0\linewidth]{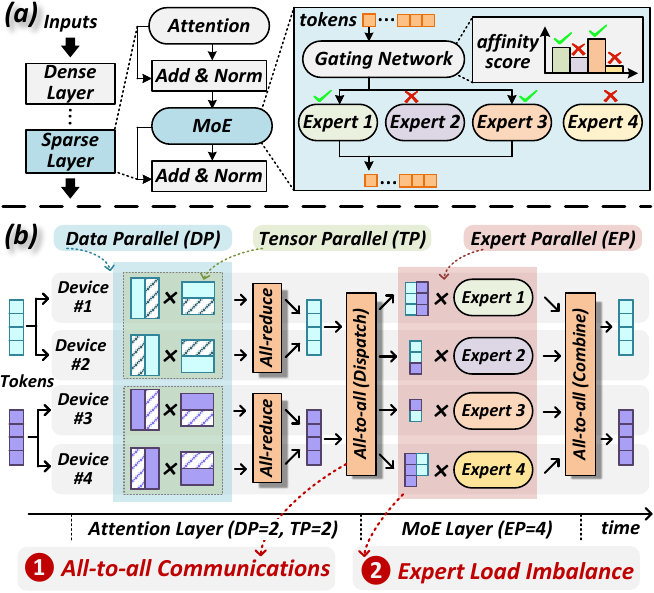}
    \caption{(a) Structure of LLM involving some sparse layers for MoE which activates 2 experts out of 4 for each token. (b) A parallelism strategy illustration with \emph{DP}=2, \emph{TP}=2 for attention layer and \emph{EP}=4 for MoE layer.}
    \label{fig:MoE_Intro}
\end{figure}

The LLMs with MoE \cite{OutrageouslyMoE} technology comprises a stack of dense and sparse blocks. Within sparse blocks, an MoE layer replaces the traditional MLP layer. Since MoE substantially reduces both computation and memory access, it has become a pivotal technology in SOTA models \cite{liu2024deepseek, qwen3report} for scaling model sizes. As illustrated in Fig. \ref{fig:MoE_Intro}(a), the MoE layer comprises a gating network and multiple expert networks, each specializing in distinct domains. The gating network selects the \emph{top-k} experts per token according to affinity scores. Tokens are then routed to their respective experts. After computation, expert outputs are weighted by the affinity scores and combined into a final output.

Fig. \ref{fig:MoE_Intro}(b) illustrates a common deployment strategy using expert parallelism (\emph{EP}) \cite{hwang2023tutel, EPS_MoE, zhai2023smartmoe} for the MoE layer, which distributes experts across devices while maintaining each expert's integrity. Although \emph{EP} enhances performance, it introduces two key problems. First, with experts distributed across devices, tokens must be dispatched to devices hosting their assigned experts and subsequently recombined on their original devices after computation. These two all-to-all communications may incur latency up to \emph{2.4×} that of computation (Fig. \ref{fig:ScaleUp}(a)), forming a major bottleneck.

Second, within each layer, certain experts stochastically attract more tokens, causing devices hosting these ``hot'' experts to experience longer computation times and severe load imbalance. While MoE models utilize auxiliary balancing losses during training \cite{fedus2022switch, shazeer2017outrageously}, inference-time load balancing remains inadequate \cite{MoE_Deploy}. Some approaches \cite{lepikhingshard} discard tokens exceeding a preset threshold, but this substantially degrades accuracy \cite{gale2023megablocks}. Thus, dynamic load balancing during inference remains essential. \textbf{In conclusion, reducing all-to-all communication overhead while alleviating expert load imbalance is crucial for efficient MoE deployment.}



\subsection{Wafer-scale Chips}

\begin{figure}[t]
    \centering
    \includegraphics[width=1.0\linewidth]{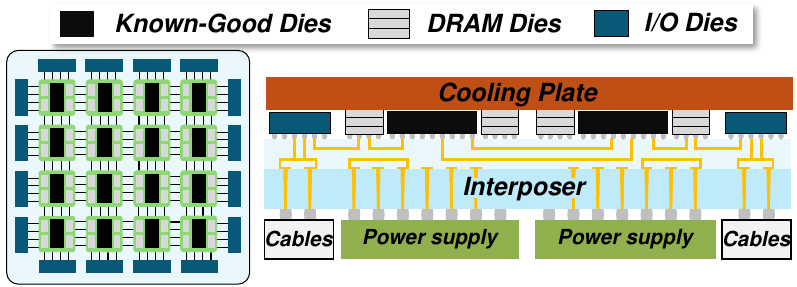}
    \caption{(a) Top view and (b) cross-section view of Wafer-scale Chips.}
    \label{fig:Wafer}
\end{figure}


Recently, benefiting from advances in Chip-on-Wafer-on-Substrate (CoWoS) technology \cite{cowos}, wafer-scale chips (WSCs) \cite{WSC-huyang, yu2025cramming,dojo,WSE-3,14336-processor} have emerged as a promising solution for hosting huge models. Fig. \ref{fig:Wafer} shows the structure of chiplet-integrated WSC: a wafer-scale interposer with metal interconnect layers is first fabricated via lithography, followed by the bonding of plenty of Known Good Dies (KGDs) onto it. Computational dies are surrounded by DRAM modules, while dedicated I/O dies integrated at the wafer periphery enable cross-wafer connectivity.


Leveraging short connection distances and an optimized interconnect hierarchy, WSCs deliver a high-bandwidth, low-latency network that spans all on-wafer and cross-wafer devices. This architecture achieves bandwidth several times higher than SOTA NVLink—up to \emph{8} TB/s—by eliminating the extensive fiber/copper cabling and router switches that contribute significantly to the cost of GPU systems. Additionally, the compact interconnects reduce I/O power consumption to as low as \emph{0.1} pJ/bit \cite{shih2025sow}, which is negligible compared to NVLink’s \emph{1.3} pJ/bit \cite{NVlink-ISSCC}. Consequently, WSCs demonstrate advantages in network performance, economic efficiency, and energy efficiency.

However, signal integrity (SI) constraints pose a dilemma in balancing link length and frequency for wafer-scale interconnects. In other words, high-bandwidth links spanning multiple dies are unachievable \cite{yang2025pd}. Thus, all industry WSCs \cite{WSE-3,dojo} adopt mesh-topology for both on-wafer and cross-wafer network, making them fundamentally different from GPU clusters in how communication should be orchestrated.

\section{Opportunity and Challenge}

WSCs demonstrate significant potential; however, due to their architectural differences from GPU clusters, directly porting optimization techniques from prior work impedes leveraging their full benefits for tangible performance gains. In this section, we first display WSCs' capability for hosting huge MoE models, then identify key challenges preventing full realization of this potential.

\subsection{The Potential of Wafer-scale Chips}

\begin{figure}[t]
    \centering
    \includegraphics[width=1.0\linewidth]{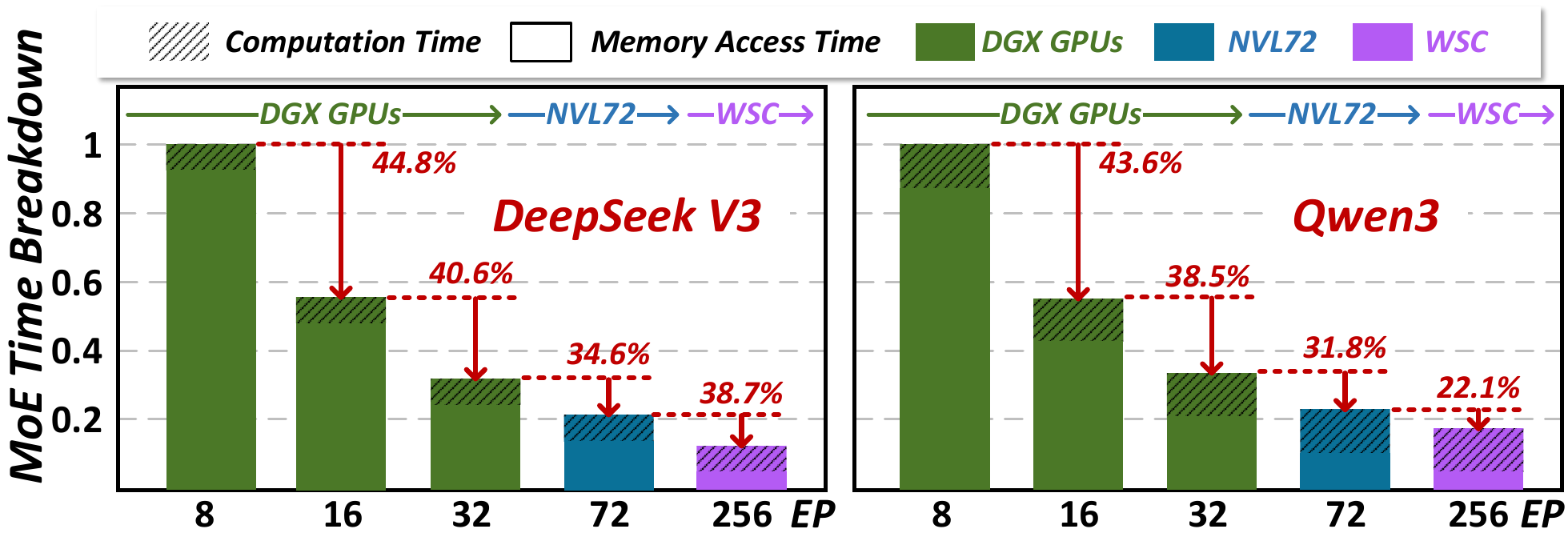}
    \caption{ The \emph{EP} that each cluster can achieve and the corresponding per-device MoE performance.}
    \label{fig:EP_scale}
\end{figure}

To optimize latency and throughput, various studies \cite{hwang2023tutel, EPS_MoE, zhai2023smartmoe} have discussed the trade-off between \emph{EP} and \emph{TP} for MoE layer. Generally, when sufficient input tokens are available, such as during the Prefill stage or in large-batch-size decoding, even with full \emph{EP}, each device can be allocated adequate tokens to maximize computational efficiency, granting \emph{EP} an advantage. When the expert size is sufficiently large to maintain computational efficiency after weight splitting, \emph{TP} becomes more advantageous. Considering that current SOTA MoE models feature numerous yet small-sized experts (e.g., \emph{256} experts with hiddenSize=\emph{2048}), \textbfIt{the EP strategy is indispensable in the MoE layer.}

The token generation phase in inference presents a severe memory bottleneck, necessitating a reduction in the number of experts per device to alleviate weight access pressure. Consequently, MoE relies on large-scale \emph{EP} to minimize experts per device—potentially down to one. As illustrated in Fig. \ref{fig:EP_scale}, increasing \emph{EP} progressively reduces memory access ratio while improving per-device performance. However, due to high-overhead all-to-all communications, the optimal \emph{EP} configuration for a cluster should match the number of devices covered by its high-performance network. For traditional DGX systems \cite{nvidiaDGX}, this corresponds to \emph{EP}=8$\sim$32. The NVL72 supernode \cite{NVL72} achieves \emph{EP}=72, yielding a \emph{35\%} performance gain. In contrast, WSCs enable \emph{EP}=256, delivering a further $39\%$ improvement and demonstrating significant potential for hosting huge MoE models. However, fully exploiting this potential requires addressing two critical challenges.

\subsection{Challenge One: Imbalanced Communication Pressure}
\label{sec:comm}

\begin{figure}[t]
    \centering
    \includegraphics[width=1.0\linewidth]{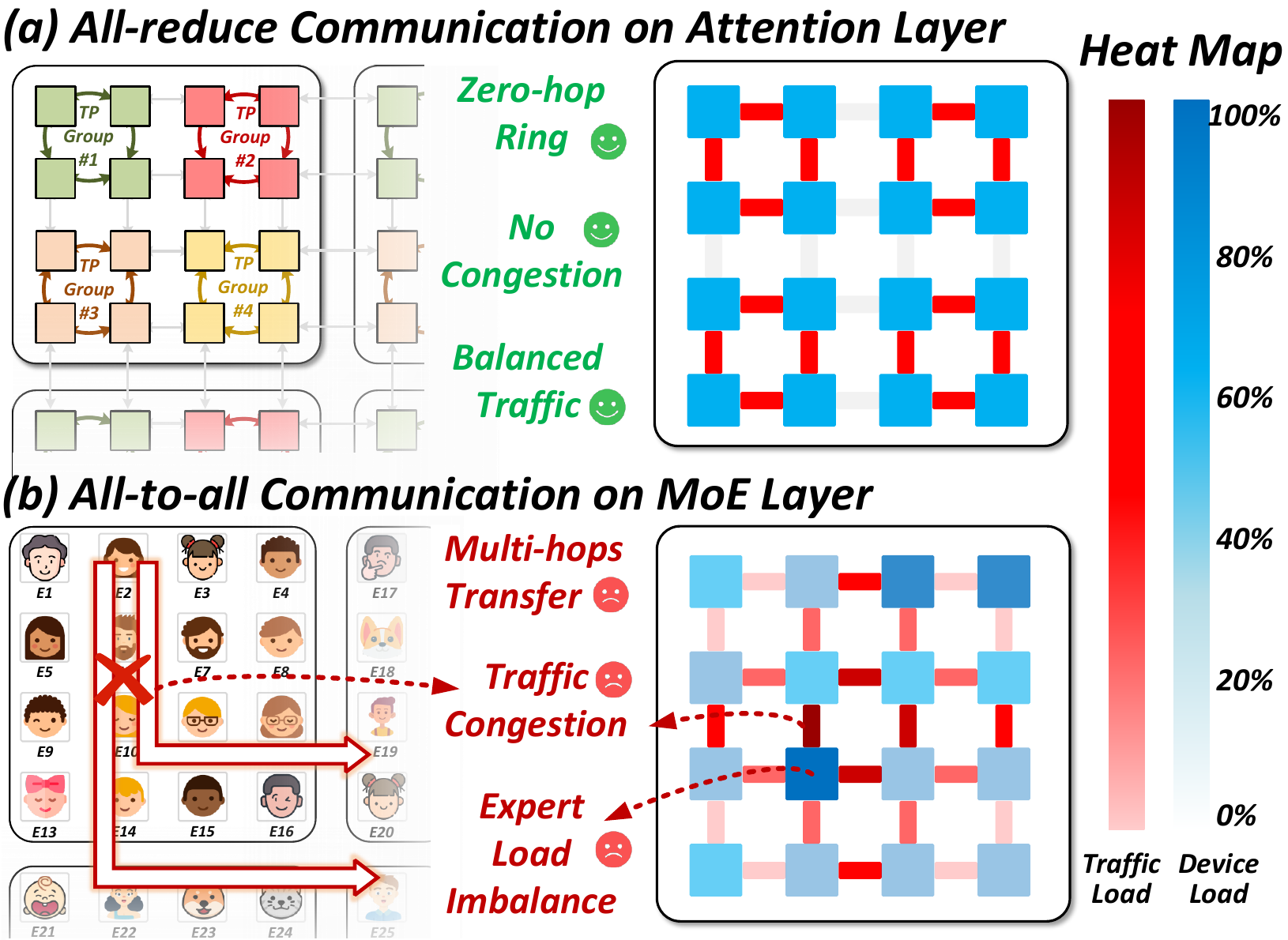}
    \caption{(a) All-reduce of attention layer with \emph{DP}=4, \emph{TP}=4. (b) All-to-all of MoE layer with \emph{EP}=16. Each face denotes an expert.}
    \label{fig:AR_A2A}
\end{figure}

\begin{figure}[t]
    \centering
    \includegraphics[width=1.0\linewidth]{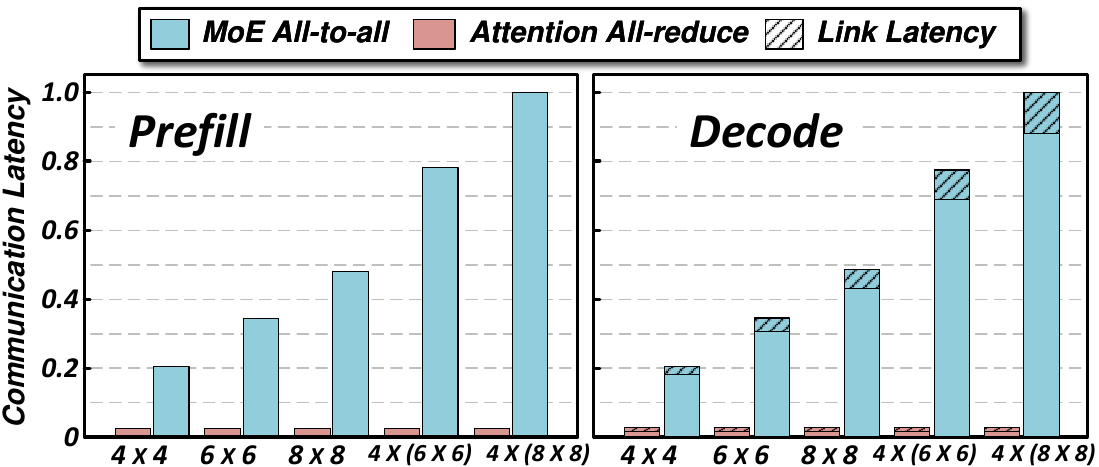}
    \caption{ Communication latency comparison between all-to-all and all-reduce; link latency during prefill stage is negligible and hence omitted.}
    \label{fig:comm_latency}
\end{figure}

\begin{figure}[t]
    \centering
    \includegraphics[width=1.0\linewidth]{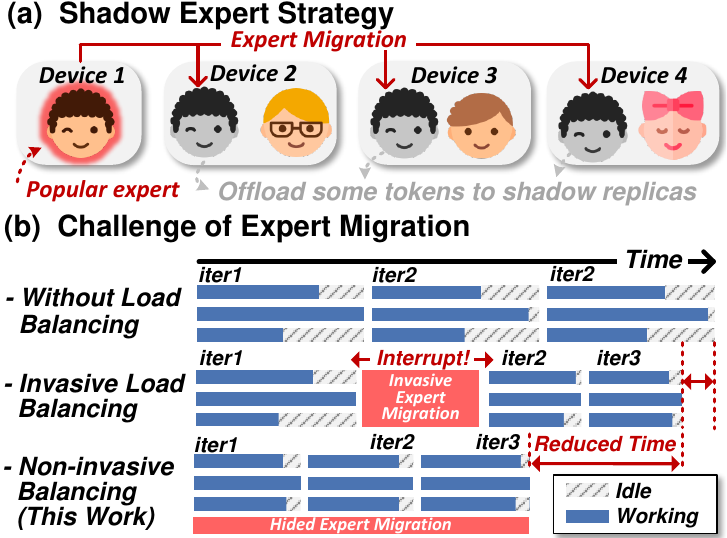}
    \caption{(a) Shadow expert load balancing strategy. (b) The Comparison of without balancing, invasive-balancing, and non-invasive balancing strategies.}
    \label{fig:Shadow}
\end{figure}

Communication latency comprises two components: data transfer time (determined by data volume and bandwidth) and link latency (governed by physical implementation and protocols) \cite{LinkLatency}. This relationship is approximated by Eq. \ref{equ:comm}, where both components are summed and multiplied by hop count. Consequently, longer distances increase communication overhead—even if individual links are fast.
\begin{equation}
\label{equ:comm}
latency = \left( \frac{volume}{bandwidth} + \mathit{link\_latency} \right) \times hops
\end{equation}

Fig. \ref{fig:AR_A2A} illustrates LLM deployment on a multi-WSC system using \emph{DP}+\emph{TP} for attention layers and \emph{EP} for MoE layers. Inputs are partitioned into segments processed by distinct \emph{TP} groups. After attention computation, ring all-reduce communication \cite{de2024swing} aggregates results within each \emph{TP} group. With hop count being one and WSC's high-performance network, this incurs minimal latency.

In contrast to the localized all-reduce, the subsequent MoE layer requires tokens to be dispatched and combined across the entire cluster via all-to-all communications. These exhibit greater complexity: tokens may reside on remote devices relative to their assigned experts, resulting in prevalent multi-hop cross-wafer transfers. Increased hop counts amplify both data transfer time and link latency, extending communication delays. Furthermore, stochastic token gating creates unpredictable point-to-point patterns that challenge orchestration, leading to concurrent transmissions congesting shared links. Expert load imbalance additionally induces traffic asymmetry that exacerbates congestion. Together, these factors make all-to-all communications count for significant time.

Fig. \ref{fig:comm_latency} compares both communications. Due to high data volume, latency is dominated by transfer time, though link latency contributes a portion in small-batch-size decoding. As WSCs scale from single \emph{4×4} platforms to multi-wafer systems, all-reduce remains trivial while all-to-all latency surges dramatically. Consequently, WSCs' high-performance network only marginally reduces all-reduce latency—yielding limited practical benefit since baseline is already low enough to overlap with computation. Conversely, WSCs' mesh topology makes all-to-all communications prohibitively expensive, establishing it as the system bottleneck. \textbf{The severe imbalance in communication pressure between all-reduce and all-to-all constitutes a critical challenge.}

\begin{figure*}[t]
    \centering
    \includegraphics[width=1.0\linewidth]{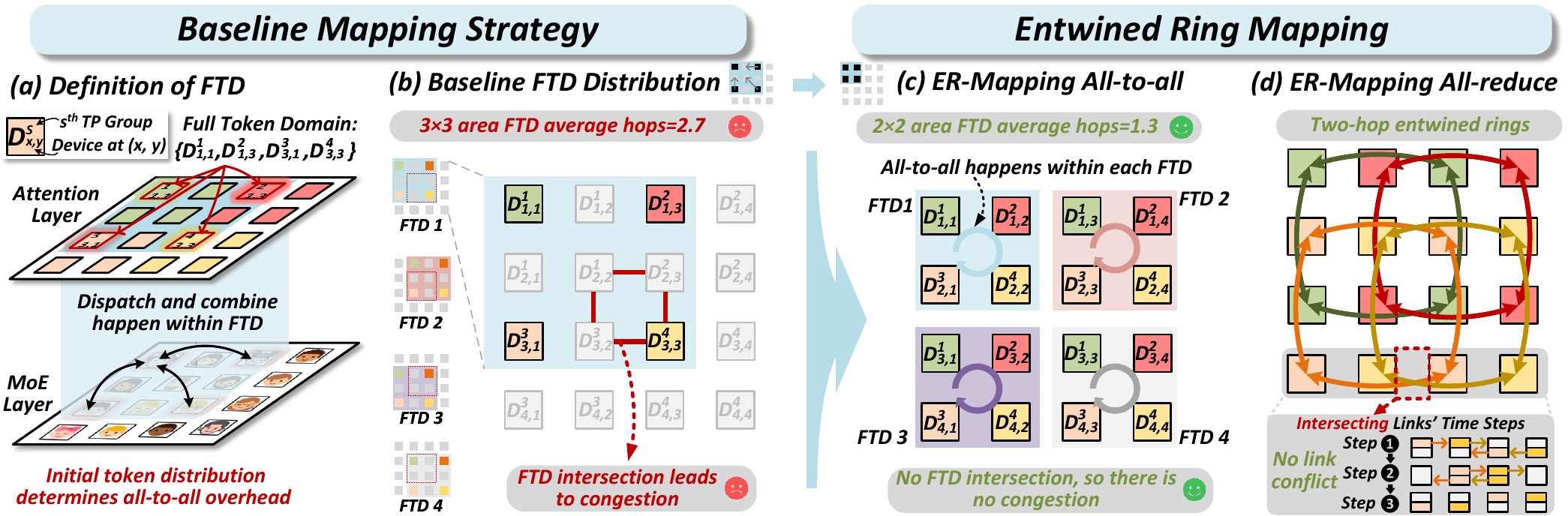}
    \caption{(a) Definition of full token domain (FTD) and interaction of attention and MoE layers. (b) FTD distribution under baseline mapping with all FTDs intersect in the center area. (c) FTD distribution under ER-Mapping which eliminates all FTD intersections. (d) Diagram of entwined ring all-reduce.}
    \label{fig:EREP}
\end{figure*}

\subsection{Challenge Two: Expert Load Balancing}
\label{sec:balance}

The load of each expert fluctuates randomly, causing load imbalance and device underutilization—an issue exacerbated under large-scale \emph{EP}. However, given the temporal continuity of load changes \cite{yun2024flex} and temporal similarity in expert selection, load prediction based on historical data is feasible. Prior works \cite{EPLB, yun2024flex, he2022fastermoe} have proposed balancing strategies for training systems. As shown in Fig. \ref{fig:Shadow}(a), devices reserve shadow slots beyond their native experts. The system predicts future popular experts using historical loads and dynamically replicates them to shadow slots on other devices. These replicas process portions of tokens for popular experts, achieving load balance.

Applying similar strategies to WSC inference systems presents unique challenges. In GPU systems, shadow slot reallocation copies expert weights from local disks via dedicated channels that avoid network contention \cite{SGLangEPLB}. However, WSC lacks on-wafer disks, forcing weight access through either wafer-edge connectors to external disks or on-wafer memory copies from devices hosting corresponding experts—both requiring high-volume multi-hop transfers across an already congested network. Furthermore, inference steps have short time spans, demanding agile balancing strategies that necessitate frequent expert migration. Yet inference serving imposes strict latency constraints. As Fig. \ref{fig:Shadow}(b) demonstrates, exposing migration on the critical path interrupts inference iterations and causes latency violations that negate balancing benefits. \textbf{Thus, ensuring balancing strategy agility without incurring latency overhead constitutes the primary challenge.}

\section{Entwined Ring Mapping}
\label{sec:ER-Mapping}

As discussed in Section \ref{sec:comm}, on WSCs, the communication latency is dominated by all-to-all while all-reduce contributes minimally. Considering that all-reduce can be overlapped with attention computation and has spare capacity, this raises a question: \textbf{can we leverage spare all-reduce capacity to alleviate all-to-all pressure?} In this section, we first explore their interaction, then propose a co-designed mapping strategy that balances communication pressure across them.

\subsection{The Definition of Full Token Domain}
\label{sec:pressure}

\begin{figure}[t]
    \centering
    \includegraphics[width=1.0\linewidth]{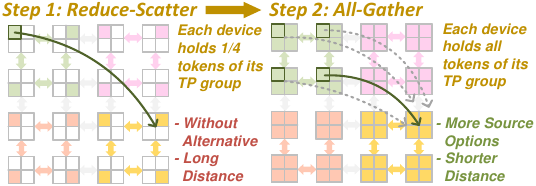}
    \caption{Benefit of retaining all-gather (AG). When a yellow device requires the highlighted portion of tokens from the green \emph{TP} group during MoE-layer all-to-all, AG provides more source options and shorter paths.}
    \label{fig:AG}
\end{figure}

As displayed in Fig. \ref{fig:AG}, all-reduce consists of a reduce-scatter followed by an all-gather (AG). While prior works \cite{magtron, pan2025fsmoe} often omit AG to reduce overhead, we find that in mesh networks, AG shortens token-fetch distances and provides routing flexibility—especially critical in all-to-all-heavy MoE workloads (evaluated in Section \ref{sec:eva_AG}). Thus, we retain AG in all-reduce first.

The mapping of \emph{TP} group in attention layers determines initial token distribution before gating, impacting MoE-layer all-to-all overhead. To figure out this interaction, we innovatively propose the \textbf{Full Token Domain (FTD)} denoting the minimal set of devices that collectively hold tokens from all \emph{TP} groups. Let $D_{x,y}^{s}$ denote the device at coordinate $(x,y)$ in the $s^{th}$ \emph{TP} group. With AG, each \emph{TP} group device holds all group tokens. As Fig. \ref{fig:EREP}(a) shows, the set $\{D_{1,1}^{1}, D_{1,3}^{2}, D_{3,1}^{3}, D_{3,3}^{4}\}$ forms an FTD by including devices from all \emph{TP} groups. Within an FTD, any device can access all required tokens, confining communication to this domain. Thus, the geometry of FTDs determines all-to-all overhead. As shown in Fig. \ref{fig:EREP}(b), using FTD, we analyze all-to-all pressure from three perspectives:

\begin{itemize}
    \item \textbf{Hops:} Assuming devices tend to access tokens from the nearest device of each \emph{TP} group, we can find four \emph{3×3} area FTDs. Ignoring load imbalance, each FTD device has uniform probability ($1/3$) of accessing tokens from any of the other three devices. Summing probability-distance products yields an ideal average of \emph{2.7} hops, implying \emph{2.7×} longer data transfer time and link latency.
    \item \textbf{Congestion:} Accurate congestion analysis is challenging as it depends on the specific routing algorithm used. We adopt an intuitive approximation: links within an FTD experience similar utilization probabilities without inherent traffic imbalance. However, under baseline mapping, all FTDs overlap at the central four devices, causing links between central devices to be shared across FTDs. This overlap induces link congestion exacerbating latency.
    \item \textbf{Imbalance:} The impact of load imbalance on congestion depends on popular expert locations. Populer experts in FTD-intersection regions intensify congestion on central links, while edge-located popular experts reduce central traffic. However, for communication-computation overlap, worst-case analysis is necessary. Thus, with FTD intersections, load imbalance increases expected latency.
\end{itemize}

\subsection{Trade-off between All-to-all and All-reduce}

Based on our analysis, all-to-all overhead is directly correlated with FTD area—larger FTDs increase average hop count and FTD intersection probability, exacerbating link congestion. To mitigate these costs, we propose Entwined Ring Mapping (ER-Mapping) minimizing the size of FTD. As shown in Fig. \ref{fig:EREP}(c)(d), ER-Mapping preserves existing parallelism configurations while co-designing attention and MoE layer mappings to balance communication pressure.

\subsubsection{\textbf{All-to-all Overhead Reduction}}

An FTD must contain devices from all \emph{TP} groups. In baseline mapping, \emph{TP} groups are spaced apart, each located in a separate corner of the mesh, which results in a large FTD area. In contrast, ER-Mapping entwines \emph{TP} groups by locating devices from different \emph{TP} groups closely together at each corner, forming compact FTDs. As Fig. \ref{fig:EREP}(c) demonstrates, the device set $\{D_{1,1}^{1}, D_{1,2}^{2}, D_{2,1}^{3}, D_{2,2}^{4}\}$ forms an independent $2\times2$ FTD. Though devices host different experts, all required tokens remain accessible within this compact domain. This configuration reduces average hops by $2\times$. It also eliminates FTD intersections, mitigating congestion, thereby reducing all-to-all latency by more than \emph{2×}.

\subsubsection{\textbf{All-reduce Latency Trade-off}}

ER-Mapping achieves all-to-all latency reduction at the trade-off of all-reduce spare capacity. As Fig. \ref{fig:EREP}(d) shows, after moving devices from different \emph{TP} groups to neighboring position, the all-reduce is transformed into four entwined two-hop rings. Packages are sent bi-directionally, step by step. Although these rings have intersecting links, the transfers are time-staggered, so there is no link conflict. Consequently, while two-hop doubles the all-reduce latency, the intersection does not worsen the latency. 

This trade-off remains advantageous. Since the base all-reduce latency is significantly shorter than all-to-all, a modest increase in all-reduce yields a substantial reduction in the more costly all-to-all. Furthermore, the increasing input sequence \cite{longseq} and chain-of-thought lengths \cite{LongCoT} in evolving LLMs dramatically expand attention computation time. This creates ample slack, allowing the longer all-reduce communication to be effectively overlapped. Consequently, the increased all-reduce latency is unlikely to degrade overall performance.

\subsubsection{\textbf{Extend to General Cases}}

\begin{figure}[t]
    \centering
    \includegraphics[width=1.0\linewidth]{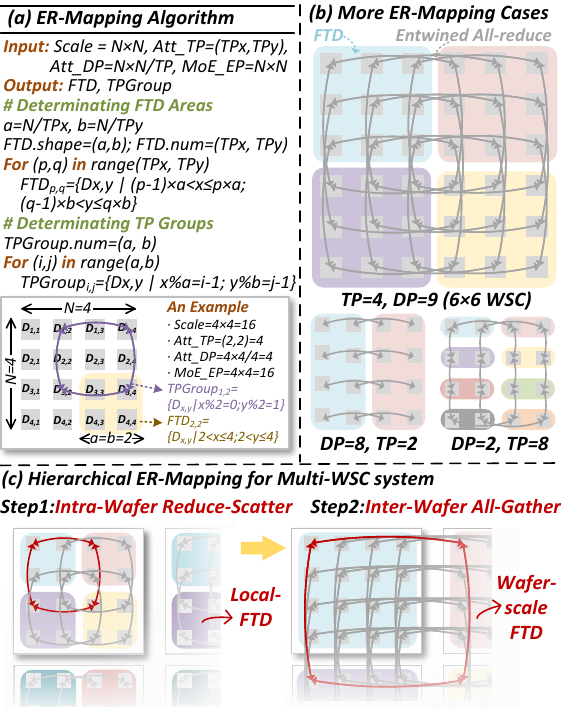}
    \caption{(a) The ER-Mapping algorithm. (b) More mapping illustrations. (c) Hierarchical ER-Mapping for Multi-WSC System.}
    \label{fig:e_EREP}
\end{figure}

Beyond the exemplary case, ER-Mapping can be readily extended to broader configurations using similar entwined-ring principles. We formalize the mapping algorithm in Fig. \ref{fig:e_EREP}(a), which takes the parallelism and WSC scale and as inputs, then returns device sets for FTDs and \emph{TP} groups. These sets define the communication domains for all-to-all and all-reduce operations, respectively. As demonstrated in Fig. \ref{fig:e_EREP}(b), ER-Mapping universally reduces the area of FTDs and eliminates their intersections, thereby balancing communication load between all-reduce and all-to-all.

\subsubsection{\textbf{Hierarchical ER-Mapping}}
\label{sec:HER}

In larger-scale systems (e.g., multi-WSC), token distribution across multiple wafers makes single entwined-ring passes prohibitively expensive. Therefore, as shown in Fig. \ref{fig:e_EREP}(c), the process splits into two hierarchical steps. First, intra-WSC reduce-scatter gathers tokens within each wafer into local FTDs—after this step, each device holds distinct token portions, enabling the entire wafer to function as a unified FTD. Second, inter-WSC all-gather aggregates tokens across wafers using these wafer-scale FTDs. Following both steps, each WSC contains tokens from all wafers, confining subsequent all-to-all exchanges within individual WSCs.

\begin{figure*}[t]
    \centering
    \includegraphics[width=1.0\linewidth]{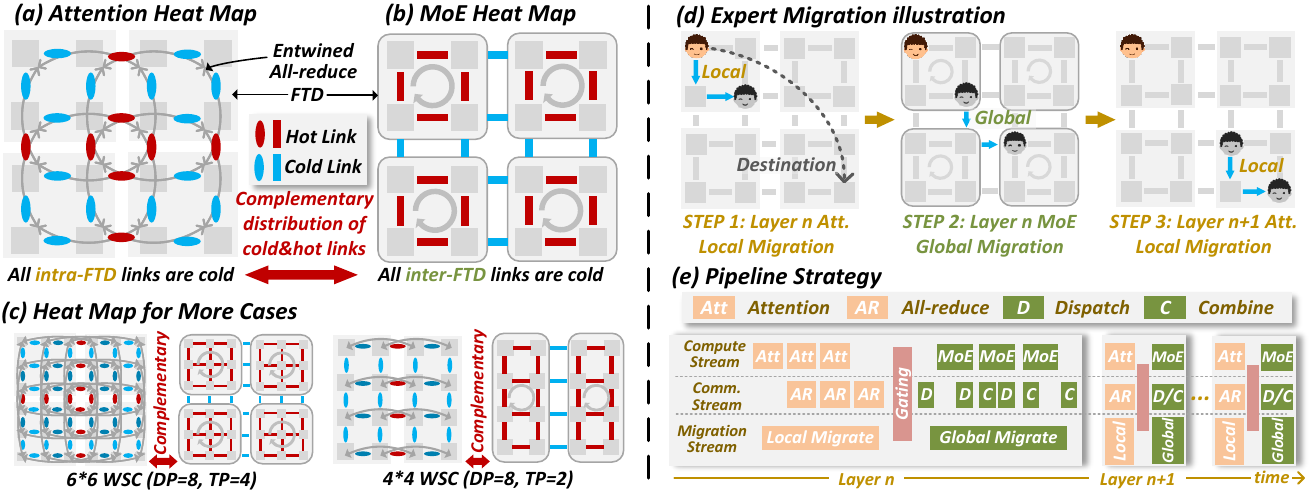}
    \caption{Traffic heatmap for (a) the all-reduce of attention layer and (b) the all-to-all of MoE layer. (c) Heatmap for more cases. The distribution of hot/cold links are complementary in all cases. (d) An illustration of expert migration. (e) The diagram inference kernels with a dedicated stream for expert migration.}
    \label{fig:migrate}
\end{figure*}

\section{Non-invasive Balancer}

As discussed in Section \ref{sec:balance}, it is critical to design an agile expert load balancing strategy while avoiding the introduction of extra latency. We therefore propose Non-invasive Balancer (NI-Balancer), which hides expert migration on the already-busy network while still delivering a satisfying and agile load balance during inference.

\subsection{Hiding Expert Migration Overhead}

During inference, the network is constantly saturated by extensive all-reduce and all-to-all traffic. To conceal expert migration overhead, identifying idle links in the busy network is essential. Fig. \ref{fig:migrate}(a) revisits the ER-Mapping design, revealing that neither the attention nor MoE layers achieve full link utilization. For all-reduce operations, links at ring intersections maintain constant activity while the others work for one cycle and then remain idle for the next cycle. Surprisingly, when flagging ``hot'' and ``cold'' links, all intra-FTD links are cold with hot links confined exclusively to FTD connection areas. This exposes spare intra-FTD bandwidth during all-reduce execution, permitting concurrent intra-FTD expert migration. Regarding MoE-layer all-to-all (Fig. \ref{fig:migrate}(b)), communication occurs strictly within non-overlapping FTDs, leaving inter-FTD connection links entirely idle and thus available for simultaneous inter-FTD migration.

Fig. \ref{fig:migrate}(c) further illustrates the heatmaps for more cases, which present similar complementary distribution of cold/hot links in these two communications. Consequently, expert migration decomposes into two operations: Local Migration (within FTDs) executed during all-reduce, and Global Migration (between FTDs) executed during all-to-all. Fig. \ref{fig:migrate}(d) exemplifies a longest-distance migration decomposed into three stages: Local → Global → Local.

\textbf{Pipelining Strategy:} To prevent communication latency from being exposed on the critical path, it is common to overlap communication with computation. In this way, as long as the communication latency is shorter than the computation time, it is acceptable. As illustrated in Fig. \ref{fig:migrate}(e), our kernel design leverages ER-Mapping’s balanced communication pressure to separately overlap all-reduce with attention computation and all-to-all with MoE computation. Inputs are split into micro-batches pipelined through computation and communication streams. In addition, there is an independent migration stream, operating when expert migration is triggered. Local and Global migrations alternately occupy idle links during each layer’s attention and MoE phases, enabling zero-overhead expert migration without disrupting regular network traffic.

\subsection{Exploiting Temporal Locality of Expert Selection} 
\label{sec:locality}

\begin{figure}[t]
    \centering
    \includegraphics[width=1.0\linewidth]{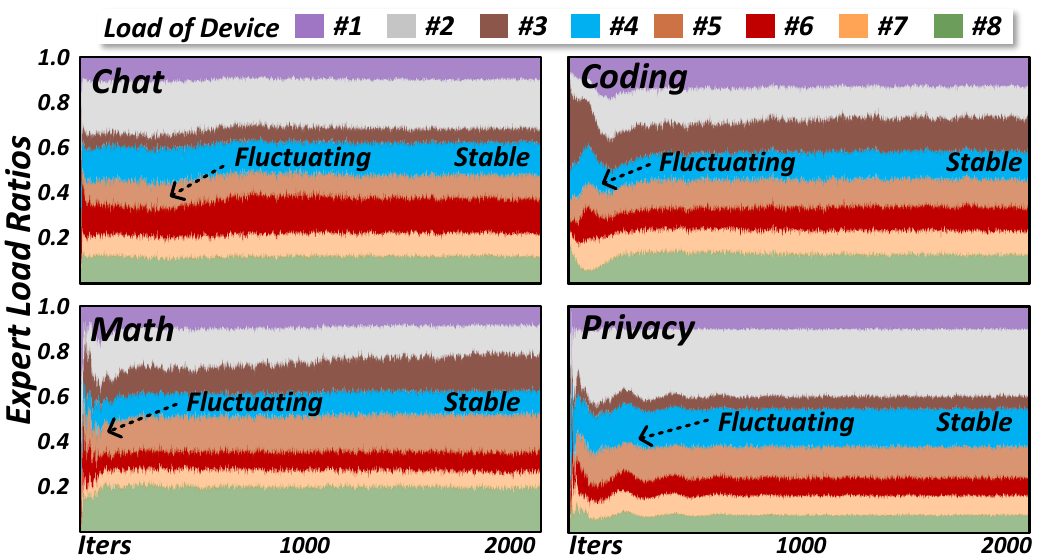}
    \caption{The expert load trace across Chat \cite{benconv}, Coding \cite{benchcoding}, Math \cite{benchmath}, and Privacy \cite{benchprivacy} scenarios. We employ Qwen3-234B with \emph{EP}=8. Each color represents one of the device load ratios. Within all scenarios, the load ratios achieve stable after a brief warm-up.}
    \label{fig:locality}
\end{figure}

During training, an auxiliary balance loss \cite{fedus2022switch, shazeer2017outrageously} encourages uniform token distribution across experts to ensure sufficient training. However, this fails to guarantee satisfactory load balance during inference \cite{MoE_Deploy}. As profiled in Fig. \ref{fig:locality}, when experts are distributed across \emph{8} devices, significant imbalance persists across all scenarios—peak device loads reach $2.9\times$ the average, causing significant device underutilization. 

However, further analysis reveals that while absolute loads remain imbalanced, device load ratios stabilize in fixed scenarios (e.g., Math-only) after initial inference iterations. This stability originates from two mechanisms: certain intrinsically popular experts consistently receive more tokens due to expert popularity bias \cite{SGLangEPLB}, and fixed scenarios persistently activate corresponding domain-specific experts across token generations \cite{yun2024flex}. This presents a balancing opportunity where expert placement can be optimized once ratios stabilize post-warmup. However, production serving encounters cyclically evolving scenario mixtures \cite{dynamollm}, where request pools gradually transition between domains, inducing slow-varying load ratios. Consequently, dynamic load balancing that continuously adapts to shifting ratios is essential.

\begin{equation}
\left\{
\begin{array}{c}
\displaystyle
\sum_{i=1}^{L} \frac{\max(\mathbf{load}_i) - \mu(\mathbf{load}_i)}{\mu(\mathbf{load}_i)} > \alpha \\[10pt]
\displaystyle
\Delta t_{\text{mig}} > \beta \quad (\beta=0\; \text{for non-invasive})
\end{array}
\right.
\label{equ:trigger}
\end{equation}

We propose a template for balancing strategy in Eq. \ref{equ:trigger}, where $\mathbf{load}_i$ represents device loads at layer $i$, and $\mu$ denotes average load. The layer imbalance degree quantifies maximum load deviation from average. Balancing triggers when the cumulative imbalance across $L$ layers exceeds threshold $\alpha$ and the time since last migration $\Delta t_{\text{mig}}$ exceeds $\beta$. Both $\alpha$ and $\beta$ are tuning parameters. For invasive balancing, token iteration interrupts to replicate popular experts to the slots of underutilized devices, with $\beta$ preventing excessive interruptions. For non-invasive balancing ($\beta=0$), migration overhead is concealed, enabling continuous fine-tuning of slots assignments.

\subsection{Enhancing Balancing Agility}

To determine migration sources and destinations, prior works \cite{EPLB, yun2024flex, he2022fastermoe} employ greedy algorithms that reassign shadow slots by directly copying the hottest expert to the coldest device. While achieving balanced load, these approaches neglect migration overhead—which can negate benefits—as selecting remote slots incurs substantially higher latency than choosing neighboring ones. Therefore, topology awareness is essential for maintaining algorithm agility.

Algorithm \ref{alg:balance} presents our topology-aware balancing strategy. We derive $Load$ from historical iteration statistics to predict expert loads. $Num_e$ denotes the number of devices hosting expert $e$ (initialized to $1$), and $Load_e/Num_e$ represents the per-device load when shared. Device $Heat_d$ is defined as the sum of $Load_e/Num_e$ for all experts on device $d$. Unlike training systems aiming for uniform token distribution, inference optimization focuses solely on reducing peak device load. Rather than targeting the globally hottest expert, we select the most popular expert on the highest-$Heat$ device as the migration source. Devices whose $Heat_d$ would not exceed the current maximum after hosting this expert constitute the $cold\_d$ set. If $cold\_d$ is empty or lacks available shadow slots, the algorithm terminates. Since any $cold\_d$ device equally reduces peak $Heat$, we select the \textbf{topologically nearest} device to minimize migration latency. After copying the source expert to the target's shadow slot, we increment $Num_e$ and update device heats. The process repeats until termination.





\begin{algorithm}[t]
\caption{\textit{Topology-aware Balancing}}\label{alg:balance}
\KwData{ $Load_e$, historical average load of $e^{th}$ expert }
\KwData{ $Num_e$, number of device hosting $e^{th}$ expert }
\KwData{ $Device_d$, experts hosted in $d^{th}$ device }
\KwData{ $Heat_d$, cumulative load of $d^{th}$ device}

$Num \leftarrow \{1\}$ ; $Heat_d \leftarrow \sum{ \frac{Load_e}{Num_e} }$ for $e \in Device_d$ \;

\While{True}
{
$hottest\_d \leftarrow max\{Heat_d\}$ \;
$src\_e \leftarrow max\{ \frac{Load_e}{Num_e} \}$ for $e \in Device_{hottest\_d}$ \;
$cold\_d \leftarrow {d}$ for $Heat_d \textless Heat_{hottest\_d}-\frac{Load_{src_e}}{Num_{src_e}}$ \;
\textbf{Break if} $cold\_d$ is empty \textbf{or} no\_slots in $cold\_d$ \;
$des\_d \leftarrow nearest\{d$ for $d \in code\_d$\} \;
Copy $src\_e$ to $des\_d$ \;
$Num_{src\_e}+=1$; Update $Device_d$ and $Heat_{d}$ \;
}

\end{algorithm}

\begin{figure*}[t]
    \centering
    \includegraphics[width=1.0\linewidth]{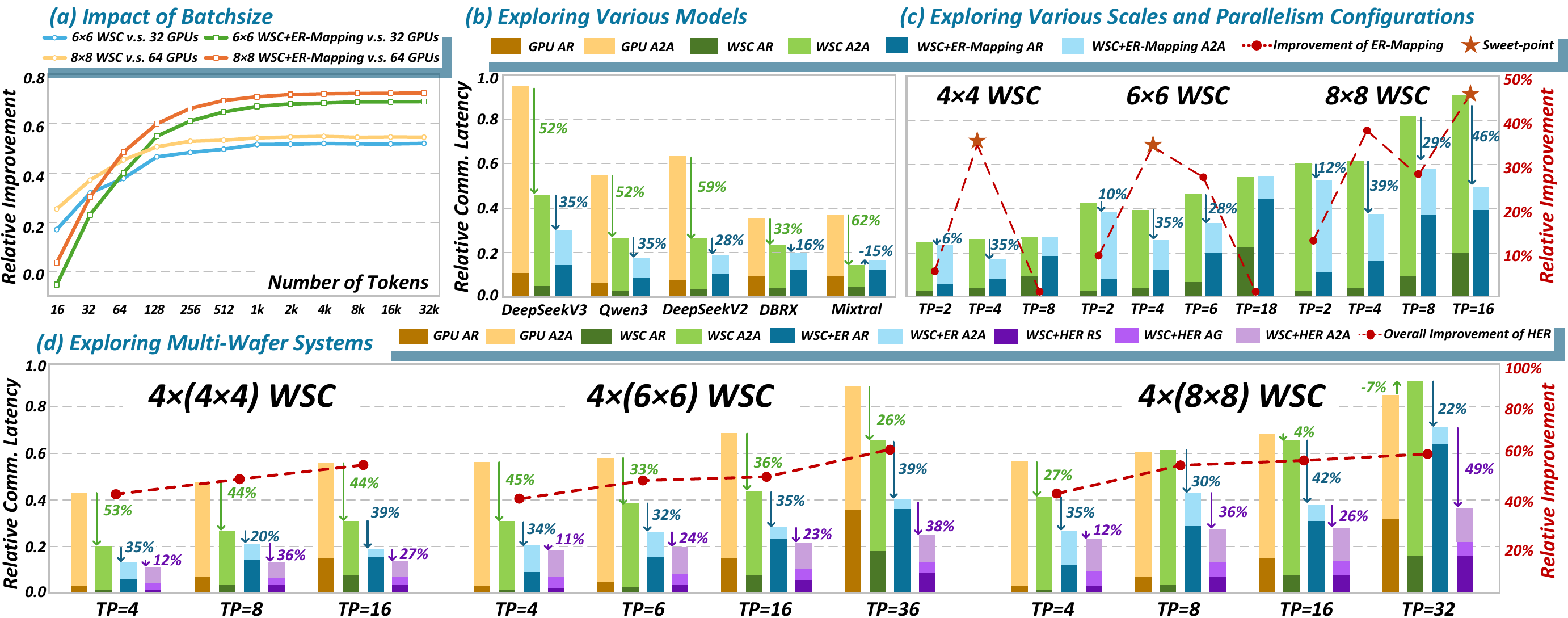}
    \caption{(a) Communication improvement of WSC over DGX under different token counts. (b) Performance of ER-Mapping under various models. (c) Exploring the impact of WSC scales and parallelism. (d) Performance of Hierarchical ER-Mapping.}
    \label{fig:exp1}
\end{figure*}

\section{Evaluation}

\subsection{Evaluation Setup}

\subsubsection{\textbf{Platform Setup}} To ensure fairness, we assume each device in the WSC is equivalent to an NVIDIA B200 GPU \cite{nvidiaDGX} capable of 2250 TFLOPS@FP16, equipped with 180GB HBM featuring 8TB/s access bandwidth. According to Tesla Dojo \cite{dojo}, the bidirectional communication bandwidth of a single die and one-border of cross-wafer bandwidth are set at 8TB/s and 9TB/s respectively. For attention layers and all communications, we employ FP16 precision, while other linear operations utilize INT8 quantization. A minimal 4×4 WSC configuration delivers 35 PFLOPS and 2.8TB memory capacities, sufficient for these huge MoE models.

\subsubsection{\textbf{Methodology}}

We employ a profile-and-simulate methodology for experimentation of both WSC and GPU clusters. The evaluator is built upon ASTRA-sim 2.0 \cite{won2023astra}—a widely recognized open-source distributed ML simulator featuring a dedicated analytical backend for network simulation. To capture dynamic inference characteristics, we profile all benchmark requests on the B200 using the vLLM framework \cite{pagedAttention}, recording input/output lengths and expert selection traces. For alignment with GPU baselines, we profile FlashInfer kernels \cite{ye2025flashinfer} across diverse input shapes on the B200, compiling a dataset of computation and memory access performance. Regarding communication performance, we first enhance ASTRA-sim's network backend with mesh-topology support. Additionally, we extend its system layer with multi-hop ring collective and point-to-point communication capabilities to support ER-Mapping and NI-Balance.

\subsubsection{\textbf{MoE Models}} To validate our optimization across various all-to-all communication overheads and parallelism configurations, as listed in Table \ref{model_tabel}, we selected SOTA MoE models with different activated expert numbers and expert sizes. This two parameters determine the magnitude of all-to-all overhead and the optimal parallel configuration, respectively.

\begin{table}[h]
\caption{Parameters of Evaluation MoE Models}
\renewcommand{\arraystretch}{1.00} 
\setlength{\tabcolsep}{4pt} 
\belowrulesep=0pt 
\aboverulesep=0pt
\centering
\begin{tabular}{l|cccc}
    
    \toprule
    \textbf{Models}  & \textbf{Size}  & \makecell{ \textbf{Layers} \\ \textbf{Sparse/Total} } & \makecell{ \textbf{Single} \\ \textbf{Expert Size} } & \makecell{ \textbf{Experts} \\ \textbf{Activated/Total} } \\ \midrule
    DeepSeek-V3 \cite{liu2024deepseek} & 671B  & 58 / 61 & 42MB & \textbf{8} / 256     \\ \midrule
    Qwen3 \cite{qwen3report}      & 235B  & 94 / 94 & 18MB & \textbf{8} / 128     \\ \midrule
    DeepSeek-V2 \cite{DeepseekV2} & 236B  & 59 / 60 & 23MB & \textbf{6} / 160     \\ \midrule
    DBRX \cite{DBRX} & 132B  & 40 / 40 & 189MB  & \textbf{4} / 16    \\ \midrule
    Mixtral-8x22B \cite{mixtral}     & 141B  & 56 / 56 & 288MB   & \textbf{2} / 8    \\ \bottomrule

\end{tabular}
\label{model_tabel}
\end{table}

\subsection{Performance of Entwined Ring Mapping}
We first explore the communication benefits of ER-Mapping across various scenarios to demonstrate its generality. To clearly isolate the sources of benefits, we initially disregard expert load imbalance. By adjusting the gating function of the MoE layer to equalize the probability of each expert being selected, we ensure balanced loads. The baseline system uses DGX B200 GPU nodes, each equipped with 8 devices, accelerated with hierarchical network communication optimization \cite{deepspeedmoe}. Both GPU and WSC employ optimizations similar to PipeMoE \cite{shi2023pipemoe} to determine the optimal pipeline stages for communication-computation fusion.




\subsubsection{\textbf{Impact of Token Count}}

To investigate how token count affects communication performance, we compare a \emph{6×6} wafer with a 4-node DGX and an \emph{8×8} wafer with an 8-node DGX. As the number of tokens per \emph{TP} group increases (Fig. \ref{fig:exp1}(a)), link latency impact diminishes, and WSC's advantage over DGX grows rapidly. Beyond \emph{256} tokens, WSC consistently outperforms DGX by \emph{54\%}, while ER-Mapping further extends this advantage to \emph{73\%}. Since token counts exceeding \emph{256} per group are achievable in both prefill and decode stages, subsequent communication experiments fix token counts at \emph{256} without distinguishing stages.

\subsubsection{\textbf{Exploring Various Models}}

We compare a \emph{6×6} WSC with a 4-node DGX to explore communication benefits under various models. Fig. \ref{fig:exp1}(b) shows that, benefited from unified high-performance network, pure WSC outperforms DGX by an average of \emph{56\%}. Additionally, in both DGX and WSC architectures, all-to-all latency is significantly higher than all-reduce, which remains minimal. ER-Mapping balances this communication pressure imbalance, substantially reducing all-to-all latency and delivering up to \emph{35\%} additional performance gains. Furthermore, since all-to-all communication overhead scales directly with the number of activated experts, ER-Mapping's benefits increase correspondingly. However, for models like Mixtral \cite{mixtral} that activate only two experts, all-to-all overhead remains relatively small while original all-reduce overhead is comparatively large. In such cases, naive ER-Mapping may fail to yield benefits.






\subsubsection{\textbf{Exploring Various Scales and Parallelism}}

We further focus on the Qwen3 to explore the impact of different configurations. As shown in Fig. \ref{fig:exp1}(c), ER-Mapping consistently outperforms the baseline, achieving improvements of up to \emph{46\%}. As \emph{TP} increases, the total token count grows, resulting in higher communication overhead. Moreover, ER-Mapping’s benefits do not scale linearly with parallelism; they are governed by the geometry of FTDs and entwined-rings, and the all-to-all/all-reduce ratio. Consequently, optimal configurations exist—for example, an \emph{8×8} WSC at \emph{TP}=16—where the topology minimizes all-to-all latency while maintaining acceptable all-reduce overhead, yielding peak acceleration.

\begin{figure}[t]
    \centering
    \includegraphics[width=1.0\linewidth]{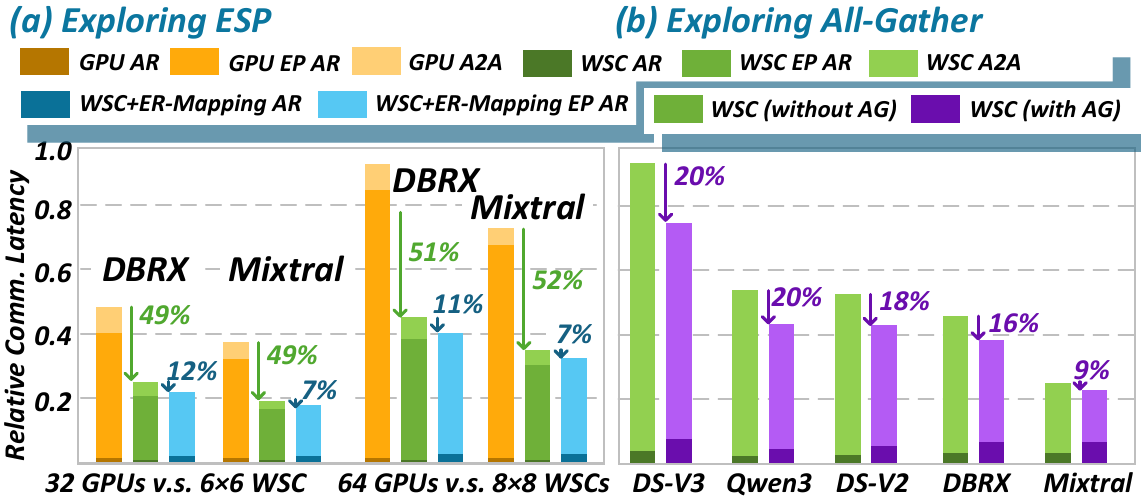}
    \caption{(a) ER-Mapping performance under ESP parallelism. (b) Justifying the retaining of all-gather.}
    \label{fig:exp1_1}
\end{figure}

\subsubsection{\textbf{Exploring Multi-WSC Systems}} 
Established in Section \ref{sec:HER}, for large-scale WSC clusters, Hierarchical ER-Mapping (HER-Mapping) is introduced to reduce multi-hop all-reduce overhead across wafers. As Fig. \ref{fig:exp1}(d) shows, HER-Mapping decouples all-reduce into two hierarchical phases—reduce-scatter and all-gather—thus further minimizing all-reduce overhead and delivering up to \emph{62\%} performance gain. Unlike pure ER-Mapping, whose performance gains vary significantly across parallelism configurations, HER-Mapping achieves consistent improvement over the baseline mapping in all cases.

\subsubsection{\textbf{Discussion for ESP Parallelism}}

Some models employ few but large-size experts (e.g., DBRX \cite{DBRX} and Mixtral \cite{mixtral}), where the substantial expert size permits further slicing. This motivates \emph{ESP} (Expert Sharding Parallelism), which further partitions individual experts across devices based on \emph{EP}. ESP necessitates all-to-all communication to gather tokens across \emph{EP} groups, followed by all-reduce operations to aggregate partial sums within \emph{EP} groups. ER-Mapping remains effective in this context: each FTD hosts several experts while distributing their slices across devices. Crucially, because all tokens across \emph{TP} groups reside within each FTD, the all-to-all communications is eliminated. As demonstrated in Fig. \ref{fig:exp1_1}(a), WSC outperforms DGX by \emph{50\%} on average, with ER-Mapping still surpassing the baseline. However, since latency is dominated by all-reduce operations within \emph{EP} groups, ER-Mapping yields only a further \emph{9\%} average improvement.

\subsubsection{\textbf{Discussion for the Retaining of All-gather}}
\label{sec:eva_AG}

As established in Section \ref{sec:pressure}, we retain the all-gather operation in the attention layer, which reduces communication distance and expands path diversity for subsequent all-to-all communications. Fig. \ref{fig:exp1_1}(b) demonstrates that while this design doubles all-reduce latency, the overhead is not significant due to the inherently low all-reduce latency. Crucially, the latency reduction from all-to-all communication offsets this cost. Consequently, after introducing AG, the performance is even improved by average \emph{17\%}, which builds the foundation for the subsequent ER-Mapping design.

\subsection{Performance of Non-invasive Balancer}

\begin{figure}[t]
    \centering
    \includegraphics[width=1.0\linewidth]{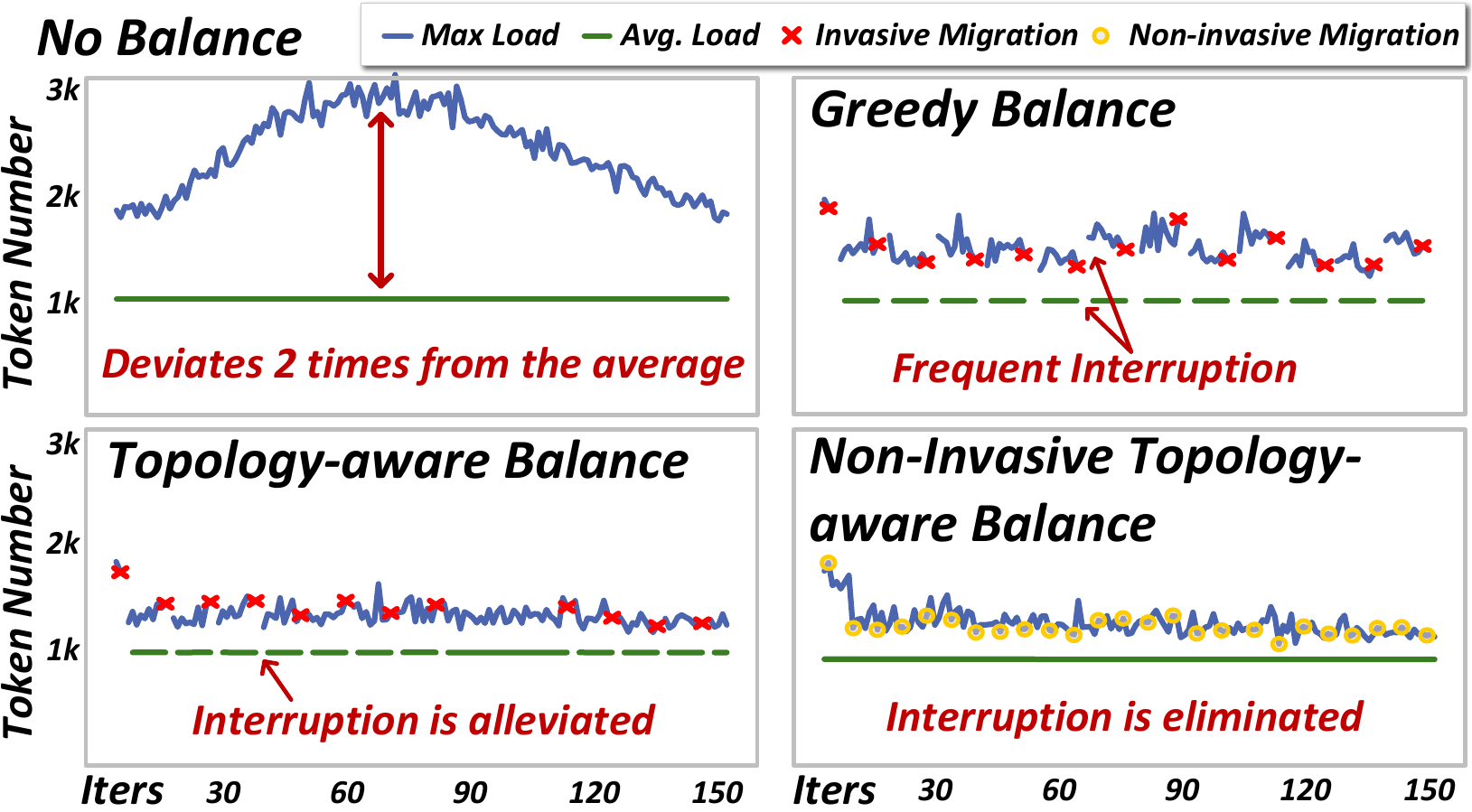}
    \caption{Run-time trace of expert loads. The green line denotes ideal average load, and the intervals on it mean the interruptions.}
    \label{fig:evluation3}
\end{figure}

\begin{figure}[t]
    \centering
    \includegraphics[width=1.0\linewidth]{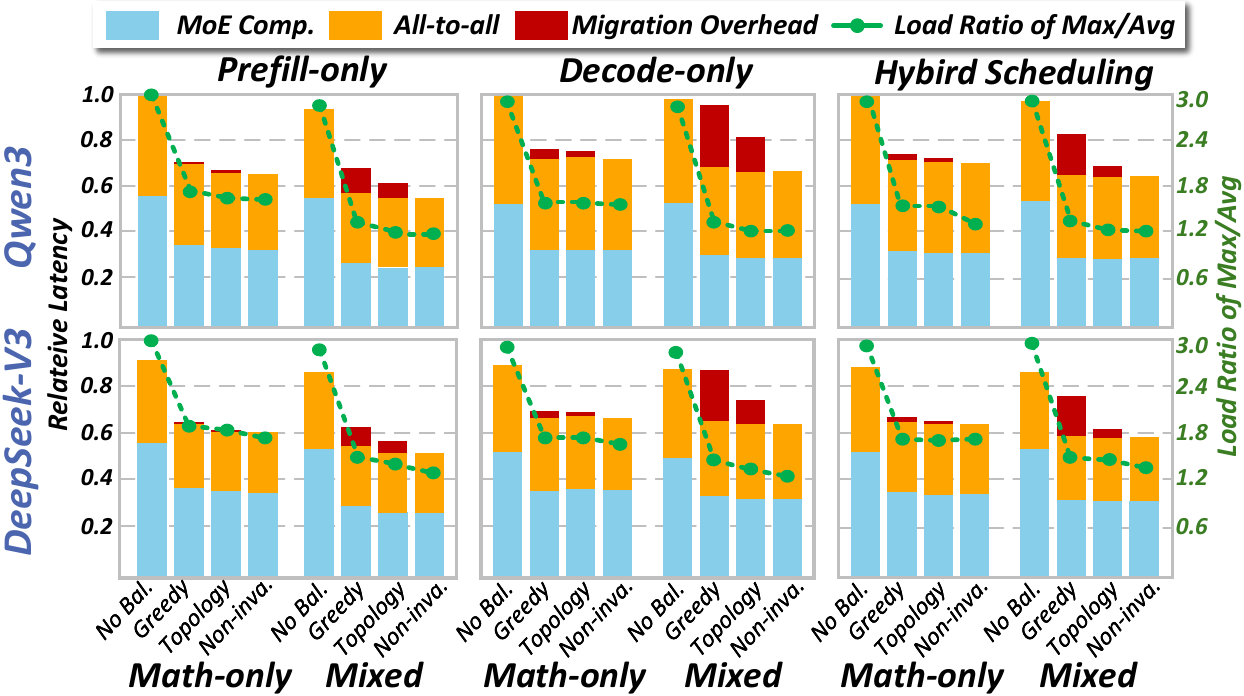}
    \caption{Comparison of different balancing strategies.}
    \label{fig:evluation4}
\end{figure}

From this section, we study the impact of load imbalance and dynamic input/output lengths. We evaluate both Disaggregated-LLM \cite{hu2024inference, mooncake, zhong2024distserve}, which separates prefill and decode on distinct platforms, and Hybird Scheduling \cite{agrawal2024taming,holmes2024deepspeed,agrawal2024mnemosyne}, which mixes them in a batch. For workloads, we leverage Evidently AI's open-source benchmark collection \cite{evident} covering four representative inference scenarios: Chat \cite{benconv}, Coding \cite{benchcoding}, Math \cite{benchmath}, and Privacy Agent \cite{benchprivacy}, where we construct single-scenario exclusively using the Math benchmark and generate mixed-scenario by integrating request arrival traces from Azure \cite{Azure} to combine all four benchmarks. The baseline greedy balancer is from EPLB \cite{EPLB}.

\subsubsection{\textbf{Run-time Load Traces}}

Fig. \ref{fig:evluation3} presents run-time traces of device loads and expert migrations. Without load balancing, the maximum load deviates by \emph{2×} from the average, causing severe imbalance and low hardware utilization. Greedy balancing reduces this deviation to approximately \emph{0.4×}. However, as an invasive method, it frequently interrupts inference iterations to perform expert migrations—triggered on average every \emph{10} iterations with overhead equivalent to \emph{2} iterations. In contrast, topology-aware balancing reduces migration distance, thereby mitigating interruptions while improving load balance. Finally, topology-aware non-invasive balancing eliminates interruption overhead entirely, allowing the balancer to remain continuously active and migrate experts whenever minor adjustments to shadow slots are required, achieving satisfactory balance.

\subsubsection{\textbf{Performance Improvement after Load Balancing}} 

We evaluate load balancing impacts across scenarios, result is displayed in \ref{fig:evluation4}. As discussed in Section \ref{sec:locality}, fixed scenarios (e.g., Math-only) stabilize load ratios after warm-up, minimizing expert migrations. However, in more common mixed scenarios, fluctuating load ratios trigger frequent migrations. Under Prefill-only, migration overhead constitutes \emph{22\%} per iteration. For Decode-only or Hybrid Scheduling, due to shorter iterations time, this surges to \emph{45\%}, potentially offsetting balancing benefits and degrading performance. Topology-aware balancing reduces migration overhead by \emph{2.6×} on average. Non-invasive balancing eliminates overhead completely, achieving optimal load balance while reducing MoE computation by up to \emph{54\%}. Additionally, balanced traffic decreases all-to-all communication time by \emph{23\%} on average.

\subsection{Ablation Study of Overall Performance}

We select NVL72 \cite{NVL72}, NVIDIA's SOTA supernode integrating \emph{72} devices with a unified high-performance network, as the baseline. Dedicated NVMe channels are adopted to hide expert migration overhead \cite{SGLangEPLB}. For WSC, we configure a multi-WSC system using four \emph{8×8} wafers (\emph{256} devices total).

\begin{figure}[t]
    \centering
    \includegraphics[width=1.0\linewidth]{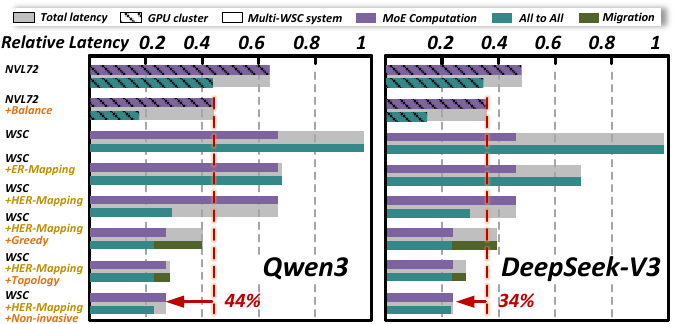}
    \caption{Multi-WSC cluster v.s. NVL72 supernode.}
    \label{fig:exp3}
\end{figure}

As illustrated in Fig. \ref{fig:exp3}, NVL72 also exhibits load imbalance. Its \emph{EP}=72 setup (multiple experts per device) leads to memory access dominating execution time, restricting load balancing gains to just \emph{26\%} computational enhancement. WSC uses \emph{EP}=256. However, rather than alleviating memory access overhead, the single-expert-per-device allocation worsens load imbalance. Moreover, the mesh topology results in all-to-all latency greatly surpassing computation time.

ER-Mapping reduces all-to-all communication by \emph{30\%}, while HER-Mapping amplifies this reduction to \emph{71\%}, eliminating communication bottlenecks. Subsequent load balancing decreases computation and communication overhead by \emph{49\%} and \emph{20\%} respectively. However, expert migration overhead on the critical path degrades overall performance. Topology-aware balancing reduces this overhead by \emph{67\%}, and non-invasive balancing eliminates it entirely. Ultimately, our optimizations remove both communication and migration bottlenecks. Compared to NVL72, WSC achieves a significantly larger \emph{EP}, delivering an average \emph{39\%} higher per-device MoE performance.

\section{Related Work}
\textbf{Communication Optimization:} Prior studies \cite{liu2024deepseek, pan2025fsmoe, pan2024parm} have optimized all-to-all communication on GPU platforms. They primarily alleviate negative impact of low-bandwidth domains through hierarchical network utilization via combined or overlapped intra-node/inter-node communications. However, WSC fundamentally differs by employing a unified high-performance network where all traffic shares homogeneous links, rendering GPU-centric approaches unsuitable. Our work uniquely exploits WSC's mesh topology to co-design all-to-all and all-reduce communications, distinguishing it from existing methods. While Lina \cite{lina} optimizes all-reduce for training gradients—inapplicable to our inference scenario—and Chimera \cite{qin2025chimera} explores communication fusion, MoE's gating network between all-reduce and all-to-all precludes such fusion.

\textbf{Topology Awareness:} To reduce the communication overhead, DeepSeek \cite{liu2024deepseek} and LocMoE \cite{li2024locmoe} proposes Node-Limited Routing, which bounds the number of nodes a token can be routed to. TA-MoE \cite{TA-MoE} incorporates topology-aware communication-cost penalties into training loss. These GPU-focused methods disregard mesh networks and inevitably constrain model capacity through routing limitations. In contrast, our approach imposes no token routing restrictions, explicitly accounts for mesh topology, and establishes a flexible platform for general MoE models without compromising capacity.

\textbf{Expert Load Balance:} Prior works \cite{fedus2022switch, shazeer2017outrageously} introduce an auxiliary loss to guarantee balanced training. However, expert loads remain imbalanced at inference time \cite{MoE_Deploy}. GShard \cite{lepikhingshard} imposes a capacity threshold and directly drop tokens that exceed it, ensuring load balance yet incurring accuracy loss \cite{yun2024flex}. Consequently, dynamic load balancing at inference remains necessary. Existing balancing strategies like EPLB \cite{EPLB}, FlexMoE \cite{yun2024flex}, and FasterMoE \cite{he2022fastermoe} typically rely on greedy algorithms, which interrupt inference and introduce significant expert migration latency. In contrast, our work explicitly accounts for migration cost, preserving algorithmic agility and hiding the migration traffic within the already busy network without causing any interruption.




 




\section*{Conclusion}

WSC presents a promising platform for hosting huge MoE models, yet their architectural distinct from conventional GPU clusters. Directly porting prior techniques hinders full utilization of WSC potential. This work introduces ER-Mapping, significantly reducing all-to-all latency on mesh networks through balanced communication. Building upon this, NI-Balancer achieves optimal load balance while concealing expert migration overhead within existing network operations. Collectively, these innovations enable WSC to deliver $39\%$ higher per-device performance compared to NVL72 supernodes.


\bibliographystyle{IEEEtranS}
\bibliography{refs}

\end{document}